\documentstyle[aps]{revtex}
\input psfig

\begin{document}

\draft
\title{Back reaction of a long range force on a Friedmann-Robertson-Walker 
background}
\author{Martina Brisudova\thanks{Electronic address: {\tt 
brisuda@phys.ufl.edu}} and William H.\ Kinney\thanks{Electronic address: {\tt
kinney@phys.ufl.edu}}}
\address{Dept. of Physics, University of Florida}
\address{P.O. Box 118440, Gainesville, FL 32611}
\date{\today}
\maketitle

\begin{abstract}
It is possible that there may exist  long-range forces in addition to 
gravity. In this paper we construct a simple model for such a force based on 
exchange of a massless scalar field and analyze its effect on the evolution of a 
homogeneous Friedmann-Robertson-Walker cosmology. The presence of such an 
interaction leads to an equation of state characterized by positive 
pressure and to resonant particle production similar to that observed in 
preheating scenarios. 
\end{abstract}


\section{Introduction}

In nature, all forces except gravity seem to be restricted in range, either 
by an intrinsic range for the interaction, as in the case of the weak force, or 
by absence of free charges, as in the case of electromagnetism. There is, 
however, no physical principle which rules out the existence of a very weak 
long-range force in addition to gravity. String theory, for example, allows for 
massless scalars which could in principle be carriers of an additional force 
and which naturally would be long range. In order to be consistent with 
experiments such as solar system tests of general 
relativity and tests of the equivalence 
principle\cite{eotovos22,will93}, such an additional force must be 
either suppressed at short distances, much weaker than gravity at all scales,  
or must be decoupled from baryonic matter altogether. The latter possibility is 
particularly interesting given that the nature of cosmological dark matter is 
currently unknown. An extremely weak self interaction in dark matter is a 
distinct possibility, and short-range self interaction has been invoked in the 
literature to explain the discrepancy between observed and simulated dark 
matter profiles in galactic 
halos\cite{spergel99,hannestad99,hogan00,burkert00,peebles00,goodman00,hannestad00,hu00,riotto00,kaplinghat00}. 
If an additional long range force existed, it could significantly modify our 
understanding of the physics of galactic scales and 
larger\cite{frieman91,gradwohl92a,gradwohl92b,kinney00}, 
but it may also 
affect the cosmology as a whole. It is the latter influence that we study here.

It has long been known that the equation of state of a thermodynamic system will 
be modified by interactions, for instance in the case of intermolecular forces 
in a gas.  On cosmological level, a change in equation of state can be expected 
to produce a change in the evolution of background cosmology. In a cosmological 
context, unlike in a more conventional thermodynamic scenario, we wish to 
address what happens if the interaction is very long (if not infinite) range 
where relativistic effects become important. Therefore the proper treatment of 
this problem requires a relativistic formulation. For this purpose, we construct 
a simple example of a long range interaction in a Lorentz invariant field 
theoretic formulation, in particular a massive scalar field $\psi$ coupled to a 
massless scalar field $\phi$  which acts as a force carrier. We study the 
evolution of these fields in the background induced by themselves, and the 
evolution of the background (i.e the scale factor).
 
In addition to the virtue of simplicity, the zero mode 
of a massive scalar provides a reasonable model for cosmological matter, since 
in the noninteracting limit a massive scalar behaves very much like a 
pressureless dust. (For recent work on the possibility that dark matter could be a scalar field, see Refs.  \cite{peebles00,hu00,riotto00,zlatev99,bento00,matos00a,matos00b,hu99,sahni99,matos99}.) An interaction will 
modify the stress-energy tensor of the cosmological matter, and thus can be 
expected to alter the evolution of the background cosmology.

The paper is organized as follows. First we introduce our model of a long range 
interaction, deliberately ignoring contact terms in order to emphasize that we 
are primarily interested in the effects of the long range component of the 
interaction. Then we study free massive scalar field in its own background, the 
behavior of the (incomplete) theory with the massless scalar exchange, and 
finally the full model included contact terms. We find the time evolution of 
fields, energy density and pressure, and the Hubble parameter in these cases. 

Our conclusions are presented in the last section.

\section{Scalar long range force}
Our model Lagrangian is:
\begin{eqnarray}
{\cal L}& = &{1\over{2}} \psi_{, \mu}\psi_{, \nu} g^{\mu \nu} +{1\over{2}} 
\phi_{, \mu}\phi_{, \nu} g^{\mu \nu} - V(\psi , \ \phi) \nonumber\\
 & =& {1\over{2}} \psi_{, \mu}\psi_{, \nu} g^{\mu \nu} - {1\over{2}} m^2 
\psi^2
+{1\over{2}} \phi_{, \mu}\phi_{, \nu} g^{\mu \nu} - \lambda \psi^2 
\phi,\label{eqbasiclagrangian}
\end{eqnarray}
where $\psi $ is the scalar field of mass $m$, $\phi$ is the massless force 
carrier, $\lambda $ is the coupling (with dimension of mass), and $\phi_{, \mu} 
\equiv \partial_{\mu} \phi$. $V(\psi , \ \phi)$ denotes the potential including 
the mass term.

In what follows, we will consider only the zero momentum states since we are 
interested in the long range effects. Obviously, the Lagrangian 
(\ref{eqbasiclagrangian}) is not a 
complete effective theory for the zero modes. 
Specifically, there are two marginal operators, $\phi^2 \psi^2$ and $\psi^4$ 
that should be included. ($\phi^4$ which is also marginal, if generated by the 
$\psi^2 \phi$ interaction, would be suppressed compared to the other two 
marginal operators by  two explicit additional powers of the coupling.) The 
marginal operators become important when the fields become large, and will be 
addressed later.

The full action of our model is
\begin{equation}
S= \int d^4x \ \sqrt{-g} {R\over{8 \pi G}} +\int d^4x \ \sqrt{-g} {\cal 
L},\label{eqaction}
\end{equation}
where $R$ is the Ricci scalar, $G =m_{PL}^{\ -2}$ is Newton's constant, and 
$\sqrt{-g}\equiv \sqrt{det(g_{\mu \nu})}.$ 
The stress energy tensor is 
\begin{eqnarray}
T_{\mu \nu} = \psi_{, \mu}\psi_{, \nu} +\phi_{, \mu}\phi_{, \nu} - g_{\mu \nu} 
{\cal L}.
\end{eqnarray}

We assume that the fields are dependent only on time, and that the background 
spacetime is a flat FRW cosmology. We use conformal 
coordinates, $ds^2 = a^2(d \eta^2 - d \vec{x} \cdot d \vec{x}) $ where $a 
\equiv 
a(\eta)$ is the scale factor, and $\eta$ is the conformal time.
Then $T_{\mu \nu}$ is diagonal, 
\begin{equation}
T^{\mu}_{\ \nu} \equiv {\rm diag}(\rho, -p,-p,-p), 
\end{equation}
where
\begin{eqnarray}
&&\rho \equiv {1 \over 2 a^2} \left[\left(\psi_{,0}\right)^2 + 
\left(\phi_{,0}\right)^2 + a^2 m^2 \psi^2 + 2 a^2 \lambda \phi \psi^2\right],\cr
&&p \equiv {1 \over 2 a^2} \left[\left(\psi_{,0}\right)^2 + 
\left(\phi_{,0}\right)^2 - a^2 m^2 \psi^2 - 2 a^2 \lambda \phi \psi^2\right].
\end{eqnarray}
The action $(\ref{eqaction})$  implies the following coupled equations of 
motion for the 
fields $\psi $, $\phi$,  and the scale factor $a$:
\begin{eqnarray}
\left(\Box + m^2 \right) \psi + 2 \lambda \phi \psi =0 \nonumber\\
\Box \phi+\lambda  \psi^2 =0 \nonumber\\
\left({{\partial_0 a}\over{a}}\right)^2 = {8 \pi \over{3 m_{PL}^2}} \rho a^2
\label{eqcoupl1}
\end{eqnarray}
where 
\begin{equation}
\Box = {1\over{\sqrt{-g}}}\partial_{\mu} \left( g^{\mu 
\nu}\sqrt{-g}\partial_{\nu} \right) = {1\over{a^2}} \partial_0^2 +2 
{\partial_0 a \over{a^3}} \partial_0.\label{eqboxdef}
\end{equation}  
$\partial_0$ denotes derivative with respect to the conformal time $\eta$.
It is convenient to introduce a {\it dimensionless} time $x$, and rescale the 
fields to absorb  $4 \pi / \left({3 m_{PL}^2}\right)$:
\begin{eqnarray}
dx &\equiv & m d\eta \nonumber\\
f  &\equiv & \sqrt{4 \pi \over{3 m_{PL}^2} }\psi , \hskip1.0truecm  
g  \equiv  \sqrt{4 \pi \over{3 m_{PL}^2} }\phi, \nonumber\\
\gamma &\equiv & \sqrt{3\over{4 \pi}} {m_{PL} \over{m^2}} \lambda.
\end{eqnarray}
The system of the coupled differential equations (\ref{eqcoupl1}) then reads
\begin{eqnarray}
\ddot{f}+ 2 {\dot{a} \over{a}} \dot{f} + a^2 f + 2 a^2 \gamma f g =0, 
\nonumber\\
\ddot{g} + 2 {\dot{a} \over{a}} \dot{g}+a^2 \gamma f^2=0, \nonumber\\
\left({\dot{a}\over{a}}\right)^2 = \tilde \rho = \dot{f}^2 + \dot{g}^2 + a^2 
f^2 + 2 \gamma a^2 f^2 g,\label{eqcoupl2}
\end{eqnarray}
Where an overdot denotes derivatives with respect to $x$, and the rescaled 
energy density $\tilde\rho$ is defined as 
\begin{equation}
\tilde\rho \equiv \left({8 \pi \over 3 m_{PL}^2 m^2}\right) \rho.
\end{equation}
We solve these equations 
numerically, subject to boundary conditions. We choose boundary conditions such 
that the force field, $g$, is only virtual (i.e. $g(x_0) = \dot{g}(x_0)=0$), 
that is, there is no $g$ radiation  because we are only interested in the 
effect of the force itself. The source field, $f$, is set to be at its free 
value initially, $f(x_0) = f_{\gamma=0}(x_0)$ and is normalized by the 
requirement that $\Omega = 1$. 

\subsection{Free scalar field in its own background}

In order to see how the long range force affects the time evolution of the 
system, it is of course necessary to first consider the case of the free field 
$f_0$, i.e. $\gamma=0$ and $g(x)=0$ at all times. 

At late times, the free field produces a background which is close, but not 
equal to, that of a pressureless dust ($\rho = w p$ with $w=0$).
It can be seen from the following: Consider time evolution of a scalar field 
$f$ not in its own background but in the background produced by a dust with 
$w=0$. (Recall that the scale factor $a \propto \eta^{2/(1+3w)}$ and $\rho 
\propto a^{-3(1+w)}$.) The equation of motion (\ref{eqcoupl2}) of the $f=f_0$ field in this 
background,
\begin{eqnarray} 
\ddot{f_0} + {4\over{x}} \dot{f_0} +x^4 f_0= 0,
\end{eqnarray}
can be solved analytically, leading to
\begin{eqnarray}
f_0(x) = N\left[{{\rm cos}\left( {x^3\over{3}}\right) \over{x^3}}+ {\rm i} {{\rm 
sin}\left( {x^3\over{3}}\right) \over{x^3}} \right],\label{eqanalyticalf}
\end{eqnarray}
where we choose the boundary conditions in such a way that $f_0^2 \equiv \vert 
f_0\vert ^2$ is monotonic, and $N$ is a constant. With the background fixed as 
$a 
\propto x^2$, consistent with our assumption of $w = 0$, the left hand side of 
the Friedmann equation is
\begin{equation}
\left({\dot a \over a}\right)^2 = {4 \over x^2}.
\end{equation}
If the solution (\ref{eqanalyticalf}) is to be self-consistent, this must be 
equal to the energy density $\tilde\rho$, given by
\begin{eqnarray}
 \tilde{\rho}= \dot{f_0}^2 + a^2 f_0^2& =& N^2\left[{1\over{x^2}} + {a^2 \over 
x^6} 
+ 9 x^{-8}\right].
\end{eqnarray}
We can normalize the field $f_0$ by fixing the scaling ambiguity in $a(x)$,
\begin{equation}
a \equiv x^2,\label{eqbackgroundscalefactor}
\end{equation}
so that 
\begin{equation}
\tilde\rho = N^2 \left[{2 \over x^2} + {9 \over x^8}\right],
\end{equation}
and setting $N = \sqrt{2}$ gives
\begin{equation}
\tilde\rho = {4 \over x^2} + O\left(x^{-8}\right) \simeq \left({\dot a \over 
a}\right)^2,\ x \gg 1.
\end{equation}
The energy density $\tilde{\rho}$ of the free scalar field $f_0$ agrees 
with the energy density of a pressureless dust up to corrections of ${\cal 
O}(x^{-8})$. This observation is useful for two reasons. First, it shows that 
for numerical purposes it is convenient to choose the initial time  $x_0 >> 1$ 
because in that case, the analytical solution (\ref{eqanalyticalf}) is close to 
the exact solution for the noninteracting field. Thus it provides a good 
boundary condition for the interacting theory if the interaction is switched on 
at the time $x_0$. Second, it allows for an independent check of the 
interacting theory via a perturbation theory with the lowest order given by the 
analytical solution $f_0$ in the $w=0$ background.

The analytical solution in the fixed $w=0$ background indicates a presence of 
small pressure, of order ${\cal O}(x^{-8})$. In the full theory, there is 
indeed a small pressure but it is oscillatory and consistent with zero. The time 
evolution of the energy density is close to $a^{-3}$. (See Figures 5 and 6.) 
Nevertheless, the scale factor grows slightly more slowly than it would in the 
strictly pressureless case. 

\subsection{The interacting case without the marginal operators}
To understand the effect of the long range exchange itself, let us first 
consider the interaction without the marginal terms $g^2 f^2$ and $f^4$.
In this case we set up a  perturbation theory to estimate time evolution of the 
fields and the scale factor. To do so, we note that the box operator 
(\ref{eqboxdef}) is easily inverted in the case where all spatial derivatives 
vanish:
\begin{eqnarray}
\Box =&& {1\over{\sqrt{-g}}}\partial_{\mu} \left( g^{\mu 
\nu}\sqrt{-g}\partial_{\nu} \right)\cr
=&& {1 \over a^4} \partial_0 \left(a^2 \partial_0\right).
\end{eqnarray}

The  fields $f$ and  $g$ can be determined order by order in $\gamma$ from their 
respective equations of motion, (\ref{eqcoupl2}). By our choice of boundary 
conditions, the $f$ field starts at zeroth order, which we approximate by the 
analytical solution $f_0$ (\ref{eqanalyticalf}), while the $g$ field starts at 
order $\gamma$.

To leading order in $\gamma$, the equations of motion for the fields are
\begin{eqnarray}
{\cal O}\left( 0 \right): \ \ \   \Box f_0 + f_0 = \left[{1 \over a^4} {d \over 
d x} \left(a^2 {d \over d x}\right) + 1 \right] f_0 = 0.
\end{eqnarray}
We take $a\left(x\right) = x^2$ (Eq. \ref{eqbackgroundscalefactor}). The field 
$f_0$ is then given by the the fixed background solution (\ref{eqanalyticalf}). 
The equation of motion for $g$ is 
\begin{equation}
{\cal O}\left( \gamma \right): \ \ \ 
\Box g_{ 1} = {1 \over a^4} {d \over d x} \left(a^2 {d \over d x}\right) g
= -\gamma f_0^2,
\end{equation}
with solution
\begin{eqnarray}
g_{1} =&& -\gamma \int_{x_0}^x{dx'  \over a^2\left(x'\right)} 
\left[\int_{x_0}^{x'}{d x'' a^4\left(x''\right) f_0^2\left(x''\right)}\right]\cr 
=&& - {2 \gamma \over{3}} \left[ \ln \left( {x\over{x_0}} \right) + 
{1\over{3}}\left( {{x_0} \over x} \right)^3 - {1\over{3}} \right].
\label{eqgperturb}
\end{eqnarray}
The corrections to $f_0$ are ${\cal O}\left( \gamma^2 \right)$, and those to 
$g_1$ are ${\cal O}\left( \gamma^3 \right)$.  To extend the perturbation 
expansion to higher order, it is necessary to include perturbations to the 
evolution of the scale factor. This can be accomplished self-consistently.

The field $g_{1}$ monotonically increases, despite the overall damping 
provided by the ever increasing scale factor. Numerical simulations confirm this 
observation. Note also  that the presence of a horizon in the FRW space 
does not save us from a singularity as we take $x_0 \rightarrow 0$. 

The behavior of the field $g$ can be simply understood. The boundary condition 
is zero field strength, $g\left(x_0\right) = \dot g\left(x_0\right) = 0$, and 
from this point the field strength gradually builds up as more and more of the 
uniform source $f$ falls within the past light cone. This continues until the 
point at which it is possible for the virtual $g$ quanta to pair-produce {\it 
real} quanta of the field $f$. This occurs when the potential energy, i.e. 
$\gamma g f^2$ becomes equal the mass term $f^2 / 2$. At this point the vacuum 
becomes unstable and  the $f$ and $g$ fields rapidly pump one another. In the 
limit that $g \gg 1 / \left(2 \gamma\right)$, the fields are increasing very 
rapidly, and we can ignore the expansion term, $\left(\dot a / a\right) \simeq 
0$ and write the equations of motion as
\begin{eqnarray}
\ddot f + a^2 (1 + 2 \gamma g) f \simeq \ddot f + 2 \gamma a^2 g f = 0, 
\nonumber\\
\ddot g + \gamma a^2 f^2 = 0, 
\end{eqnarray}
from which we can derive a simple relation for the field strengths,
\begin{equation}
f \ddot f = 2 g \ddot g.
\end{equation}
Therefore the fields grow without bound as $f = \sqrt{2} \vert g \vert$, a 
behavior which 
is observed in the numerical solution. (Remarkably, our simple perturbation 
theory underestimates the field $g$ at the onset of the instability by only 20\%.) This relationship between the field strengths has a simple interpretation 
in terms of particle number. Since the number densities are proportional to the 
square of the field strength, $n_f \propto f^2$, an interaction of the form $f^2 
g$ produces twice as many quanta of $f$ as quanta of $g$. 
Figure 1 shows the the exact numerical solutions for the fields $f$ 
and 
$g$ compared to the free analytical solution $f_0$ (\ref{eqanalyticalf}), and 
the 
lowest order 
perturbative expression   (\ref{eqgperturb}), respectively, for $\gamma 
=0.8$. The 
approximate solutions agree very well with the numerical results until very 
close to the 
onset of the instability. The plot of $f$ shows particularly distinctly the 
sharp rise of the $f$ field at the onset of the instability.

One would expect the approximate perturbative solutions $f_0$ and $g_1$ 
(\ref{eqanalyticalf}), (\ref{eqgperturb}) to break down in 
the limit of strong coupling, $\gamma > 1$. In practice, the perturbative 
expressions work very well not only for weak coupling $\gamma <1$, but 
for relatively large $\gamma$ as well. Figure 
3 shows the fields $f$ and $g$ compared to the approximate 
solutions (\ref{eqanalyticalf}) and (\ref{eqgperturb}) for $\gamma=5$. Note that 
at the onset of instability the field $g$ is 
again underestimated by only 20\% (as for $\gamma =0.8$). The onset of 
instability is seen in both $f$ and $g$. 

This analysis, however, neglects one very important effect: the field $g$ is not 
only coupled to the zero mode of $f$, but it is also coupled to modes with 
nonzero momentum. For a mode $f_\kappa$ with dimensionless momentum $\kappa 
\equiv k / m$, the equation of motion is:
\begin{equation}
\ddot f_\kappa + 2 \left({\dot a \over a}\right) \dot f_\kappa + a^2 \left(1 + 
\kappa^2 + 2 \gamma g\right) f_\kappa^2 = 0.
\end{equation}
The field $g$ can pair produce modes $f_\kappa$ when
\begin{equation}
2 \gamma g > 1 + \kappa^2,
\end{equation}
so that as $g$ grows, more and more phase space is populated through pair 
production. A proper treatment of particle production in this circumstance
should invoke non-equilibrium dynamics, and is beyond the scope of our simple 
calculation. Similar systems have been studied in the literature 
\cite{boyanovsky94,boyanovsky97}.

When the fields become large, however,  the long range force is no longer the 
dominant factor. The contact interactions $g^2 f^2$ and $f^4$ become important.
The instability is unphysical. 
 
\subsection{Including the marginal operators}
 To stabilize the potential,
 \begin{equation}
 V(g , \ f) = {1\over{2}} f^2 + \gamma g f^2,
 \label{eqpotential1}
 \end{equation}
  that is to make it bound from below, we now add the even marginal operators 
mentioned earlier.  For our purposes, their coefficients are {\it a priori} 
unknown, 
because we use this model as an effective theory. In principle, the 
coefficients 
could be derived by integrating out degrees of freedom in the process of 
deriving the effective theory from the full theory at a different scale. 
 
There are three marginal and even operators:

{$\bullet \  {\boldmath g^2 f^2}$}

\noindent
This operator by itself is sufficient to make the potential bound from below, 
providing its coupling is large enough. 
The potential (\ref{eqpotential1}) with this operator can be rewritten as 
follows:
\begin{eqnarray}
V(g , \ f) \longrightarrow {1\over{2}} f^2 + \gamma (g f^2 + c g^2 f^2)
= \gamma c (g  + {1\over{2 c}})^2 f^2 + {1\over{2}} (1- {\gamma\over{2 c}}) f^2,
\end{eqnarray}
which shows that ${\gamma/({2 c})}$ has to be less than one, or else the 
minimum of the potential would be negative infinity at $g = -{1/({2 c})}$, 
and 
$f^2 \rightarrow \infty$. In other words, the interaction decreases the mass 
of the $f$. If the marginal operator is too weak the {\it effective} mass of $f$ 
becomes tachyonic as the field $g$ flows to its minimum.

{$\bullet \  {\boldmath g^4, \  f^4}$}

\noindent
Neither one of these operators alone can stabilize the potential. $f^4$ allows 
for potential to be unbound for $g \rightarrow -\infty$, $f$ finite; and 
$g^4$ allows for $g$ finite, $\gamma g <-1,$ and  $f \rightarrow \infty$.
The sum of these two operators would make the potential to have a finite 
minimum.
However, we have argued that the operator $g^4$, if generated by a $g f^2$ 
interaction, would have to be suppressed compared to the other two marginal 
operators $g^2 f^2$ and $f^4$ by two additional powers of coupling of the $g 
f^2$ interaction. 
($g^2 f^2$ and $f^4$ can be expected to be of the same order.)
For this reason, and in order to reduce number of parameters, we use only 
$g^2 f^2$ and $f^4$. Further, for simplicity we assume that both these 
operators 
have the same coefficients. (This is a reasonable assumption if these operators 
arise from exchange of high energy $f$ or $g$, respectively.) The qualitative 
features of the results are the same for unequal couplings as well.

The potential including the marginal operators is
\begin{eqnarray}
V_{\rm full}(g , \ f)= {1\over{2}} f^2 + \gamma (g f^2 + c g^2 f^2 +c f^4)
= \gamma c (g  + {1\over{2 c}})^2 f^2 + \gamma c f^2 (f^2 - \tilde{f}^2),
\label{eqpotential}
\end{eqnarray}
where 
\begin{equation}
\tilde{f}^2 =-{1 \over {2 \gamma c}} (1- {\gamma/{2 c}}).
\end{equation}
The minimum of this potential is always finite, regardless of the value of $\gamma$ 
and $c$. If $\tilde{f}^2 \leq 0$, or equivalently, $\gamma < 2 c$, the absolute 
minimum of the potential is zero, at $g=-1/(2 c)$ and $f=0$. The case  
$\tilde{f}^2 > 0$ corresponds to a ``broken" phase, similar to a ``Mexican hat'' 
potential. The potential still has a local extremum equal zero at $g=-1/(2 c)$ 
and $f=0$, but the absolute minimum is at $f = \pm \tilde{f}$, $g=-1/(2 c)$  and 
it is negative, $V_{min} = - \gamma c \tilde{f}^4$. The energy density can 
become negative. To make the theory consistent, one must add a constant to the 
potential so that the energy density  is positive definite.

The numerical results that we present here were generated using the potential 
(\ref{eqpotential}) in the unbroken phase.  Even though the broken phase might 
seem like a better choice to represent the effect of the long range force 
because the marginal operators can be made arbitrarily weak, we find that it is 
less convenient for numerical reasons. In addition, we wish to avoid the 
presence of inflationary solutions. The results that we show are 
representative of a number of calculations with various couplings.

The evolution of the fields is at first identical to the case with no marginal 
operators because the long range component is dominating the physics (see 
Figure 3). The onset of the rapid growth of the fields in the presence of 
the marginal operators, Eq. (\ref{eqpotential}), is delayed, compared to 
the onset of instability discussed in previous section, Eq. 
(\ref{eqpotential1}). Another difference between the two cases is that with 
the marginal operators, the field $f$ reaches its maximum and then starts 
oscillating. This is reminiscent of parametric particle production in 
preheating\cite{kofman97}, and in fact similar. The difference is that in our 
case it is the 
energy stored in the massless messenger field that causes the peaks in the 
massive field $f$. The parameters in the calculation shown in Figure 3  
are chosen such that the field $f$ has a very small effective mass when $g$ 
reaches $-1/(2 c)$; however the peaks in the field $f$ persist even if the mass 
is not close to zero. They do become wider and less steep with the increasing 
effective mass.

The plots also show the  perturbative approximations for the fields $f$ and $g$. 
The $g$ field is well approximated by the perturbative expression up until the 
growth of $g$ slows down and eventually ceases due to the marginal operators. 
Subsequently, $g$ 
is seen to 
oscillate around the value  $-1/(2 c)$. The peaks in $f$ field occur when $g = 
-1/(2 c)$, that is the point in the potential valley when the effective $f$ mass 
is minimal. The field $f$ is also close to its free value until the onset of the 
instability. After that, in addition to rapid oscillations of the amplitude that 
are absent in the free case, the fully interacting field  decreases more slowly 
than the free field.

Next we show energy density and pressure. The long range force gives rise to a 
positive pressure, as can be expected from an attractive interaction. In the 
absence of marginal operators, the instability occurs 
when the pressure becomes equal to the energy density. At that point, the energy 
density drops to zero while pressure continues to grow and the theory becomes 
non-causal, $p > \rho$. (Indeed, the $f$ field is effectively tachyonic.) This 
is just an indication that our ``effective theory'' for the long range force, 
consisting of just the relevant operator, is incomplete. When the  marginal 
operators are included, the 
pressure builds up slower, and when it becomes equal to the energy density, it 
decreases and starts oscillating, following the  behavior of the $f$ field.
The energy density decreases monotonically, but it has saddle points.

Although the behavior of the interacting matter differs markedly 
from the free theory, the effect on Hubble parameter is at most about 20\% 
(Figure 5). This is partially because we start at time $x_0$ when the scale 
factor is already large.

The Hubble parameter for the free field is practically indistinguishable from 
the Hubble parameter due to a pressureless dust. In the case of the interacting 
theory with no marginal operators, the Hubble parameter is close to that of the 
free field, and then drops (to zero) when the instability occurs. This could be 
expected since the energy density is dropping also. With the marginal operators, 
the Hubble parameter is still monotonic, but it has saddle points at times when 
the amplitude of $f$ field peaks and $g$ field is equal $-1/(2c)$, coinciding 
with the saddle points of the energy density. The rapid oscillations of $f$ are 
too fast to affect the Hubble parameter.At the first saddle point, the 
difference between the Hubble parameter of the free vs. interacting field is 
about 20\%.

\section{Conclusion}

In this paper we analyze a simple model for a long-range interaction in 
cosmological matter, based on exchange of a massless scalar ``messenger'' field 
$\phi$. The boundary conditions are set such that the interaction is turned on 
at a finite time $\eta_0$ in a background consisting of the pressureless zero 
mode of a massive scalar field $\psi$ in a flat Friedmann-Robertson-Walker 
spacetime. We assume an interaction of the form $\phi \psi^2$. This coupling 
creates a {\it negative} interaction potential for the matter field, which, if 
not regulated by marginal operators in the Lagrangian, results in an unstable 
vacuum. The instability has a simple physical interpretation: the messenger 
field gradually builds in strength until the energy density in the field is 
large enough to pair-produce real matter particles. The particle production 
results in further strengthening of the messenger field and feedback results in 
strong pumping of both the matter and the messenger field. Since the potential 
is unbounded from below, this feedback continues indefinitely (Figs. 1 and 2). 
We observe that the matter field rapidly becomes tachyonic and the theory is 
non-causal, with pressure exceeding the energy density. With the addition of 
marginal operators of the form $\phi^2 \psi^2$, $\phi^4$ and $\psi^4$, the 
potential is bounded from below and the theory is well-defined. The early-time 
behavior of the theory is the same regardless of the presence of the marginal 
operators, but at late times the messenger field enters an oscillatory phase, 
resulting in resonant particle production in the matter field, similar to that 
seen in preheating scenarios (Fig. 3). We analyze the effect of the interaction 
on the equation of state of the cosmological matter and find that it results in 
the generation of an oscillating positive pressure term, with a maximum at $p = 
\rho$ (Figs. 5 and 6). This has two effects on the expansion. First, it results 
in an overall slowing of the expansion rate by a factor of about 20\%. Second, 
the oscillations in the equation of state result in a deviation of the Hubble 
parameter from a simple power-law dependence on the conformal time (Fig 4).

The interaction we study here is mediated by a scalar and thus represents an 
{\it attractive} force. This is the simplest case, but potentially not the most 
interesting. A repulsive interaction, for instance one mediated by a 
vector field, could potentially provide a model for the mysterious 
``quintessence'' responsible for the observed accelerating expansion of the 
universe. This is the subject of continuing study.

\section*{Acknowledgments}

We would like to thank Dallas Kennedy and Richard Woodard for helpful 
conversations. This work was supported in part by U.S. D.O.E grant 
DE-FG02-97ER-41029.

\vfill\eject

\begin{figure}
\psfig{figure=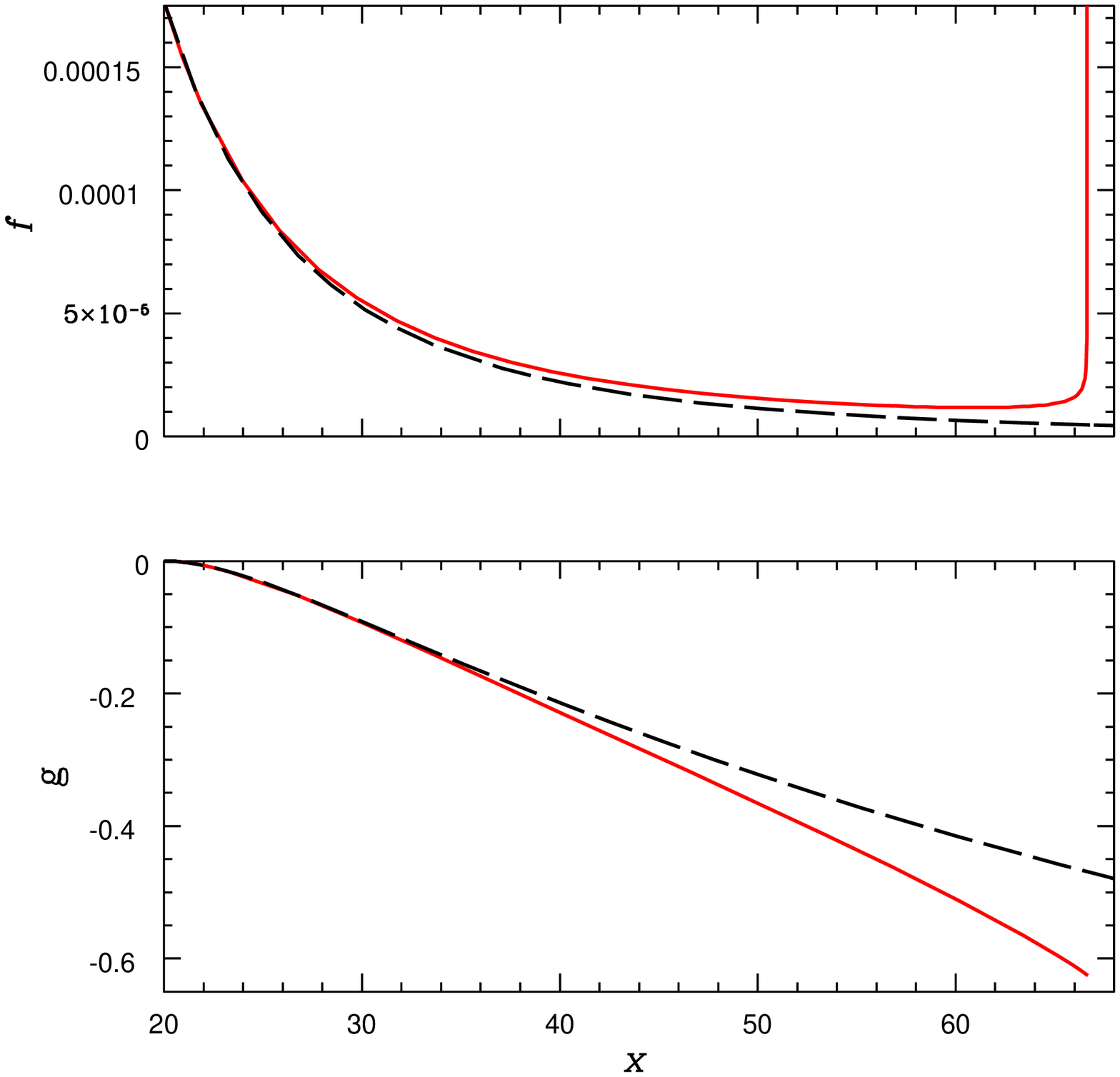}
\bigskip
\bigskip
\bigskip
\bigskip
\bigskip
\bigskip
\caption{Perturbative (black, dashed) and numerical (red, solid) solutions for 
the 
fields $f$ and $g$ with $\gamma = 0.8$.}
\end{figure}

\begin{figure}
\psfig{figure=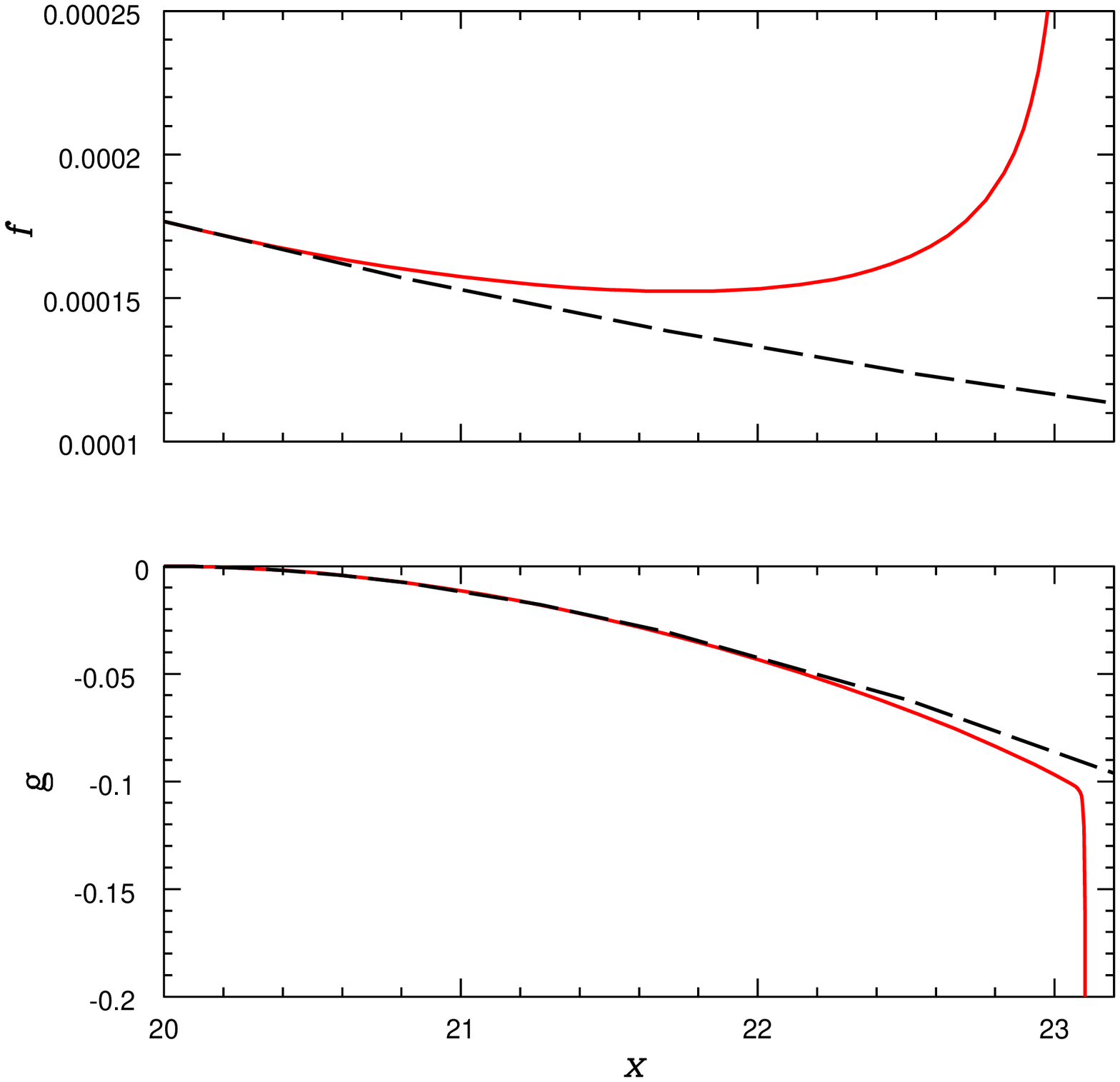}
\bigskip
\bigskip
\bigskip
\bigskip
\bigskip
\bigskip
\caption{Perturbative (black, dashed) and numerical (red, solid) solutions for 
the 
fields $f$ and $g$ with $\gamma = 5$.}
\end{figure}

\begin{figure}
\psfig{figure=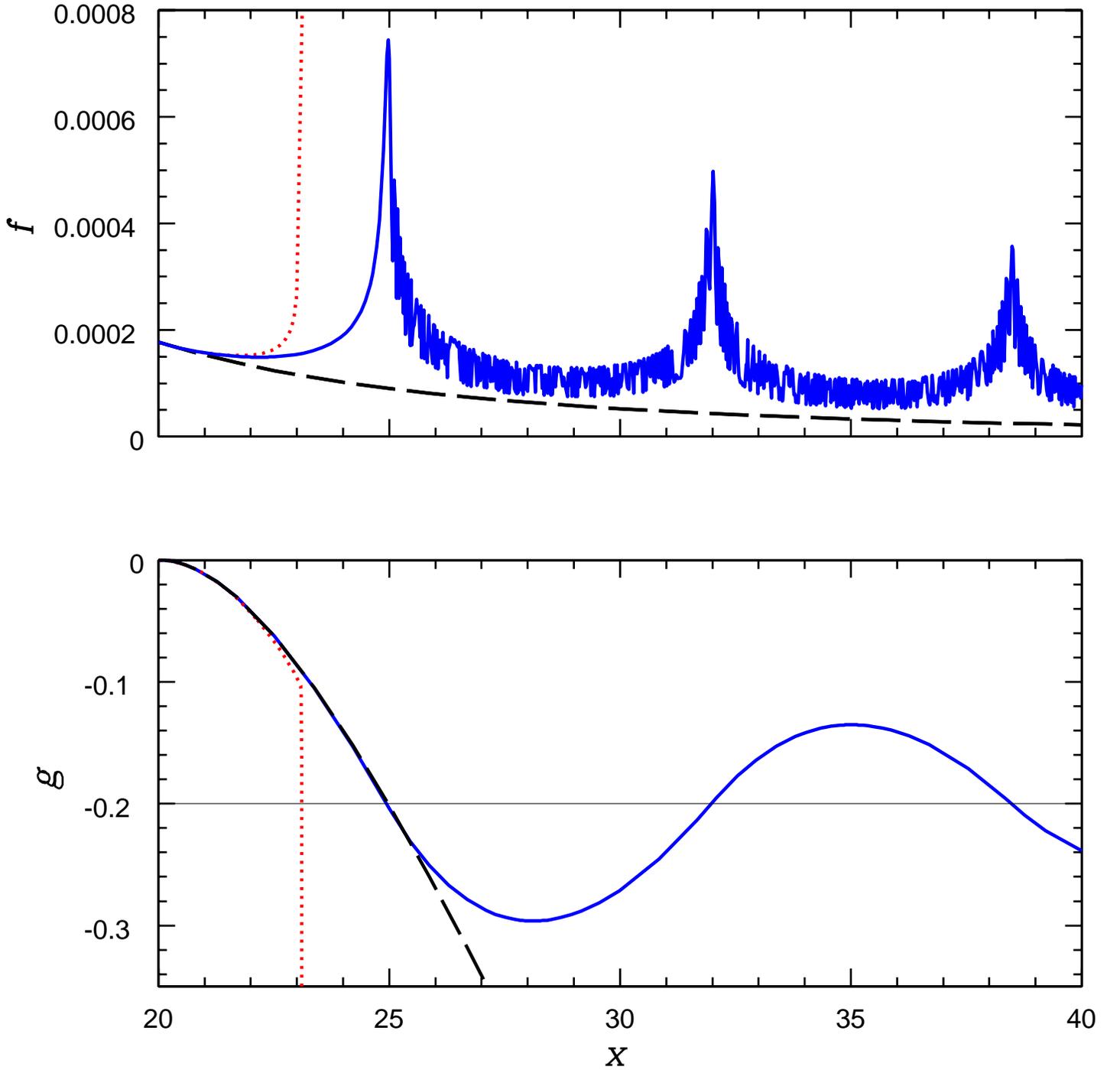}
\bigskip
\bigskip
\bigskip
\bigskip
\bigskip
\bigskip
\caption{Solutions for the fields $f$ and $g$ with $\gamma = 5$ for the case 
including marginal operators. The solid (blue) line shows the numerical solution 
for the fields. The perturbative solution is shown as a dashed (black) line, and 
the case without marginal operators is shown as a dotted (red) line. Note in 
particular the resonant amplification of $f$ when the messenger field passes 
through $g = -0.2$.}
\end{figure}

\begin{figure}
\psfig{figure=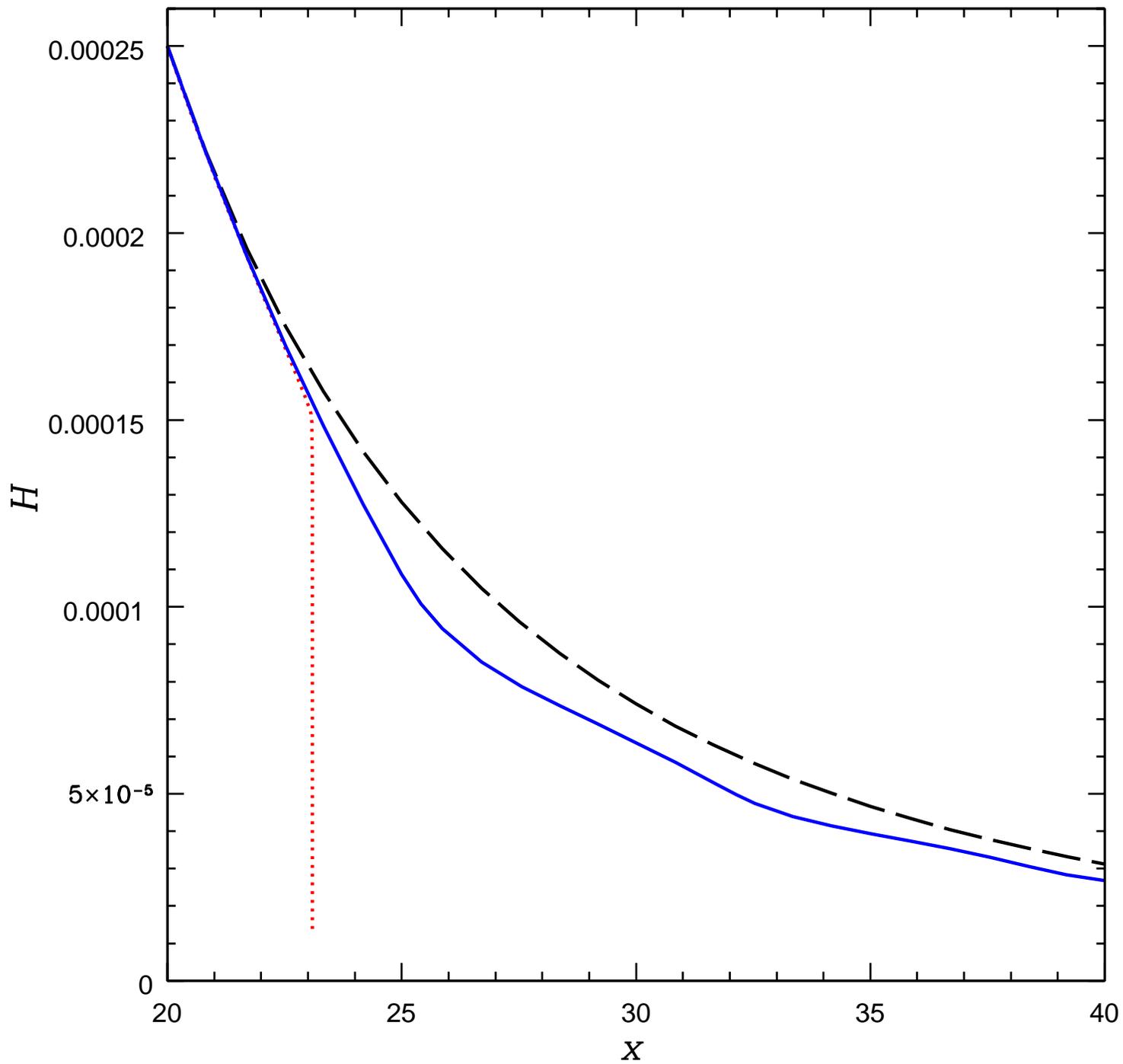}
\bigskip
\bigskip
\bigskip
\bigskip
\bigskip
\bigskip
\caption{Hubble parameter as a function of time. The black (dashed) line 
represents the free field case. The red (dotted) line is the case without 
marginal operators included. The blue (solid) line is the case with marginal 
operators included.}
\end{figure}

\begin{figure}
\psfig{figure=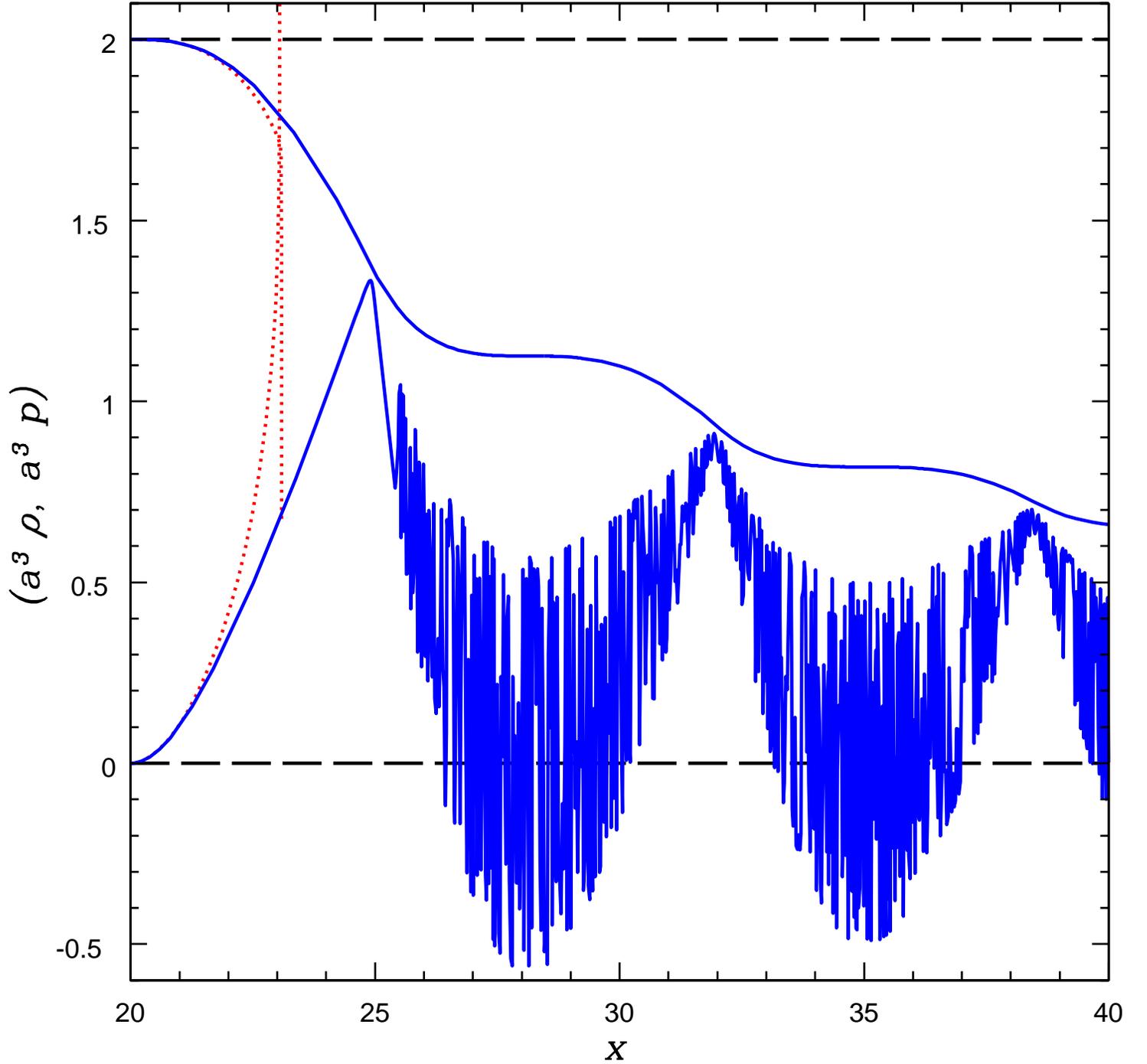}
\bigskip
\bigskip
\bigskip
\bigskip
\bigskip
\bigskip
\caption{Energy density and pressure for the case $\gamma = 5$. The black 
(dashed) line is the free field case: $a^3 \rho = 2$ is the top line, and $a^3 p 
= 0$ is the bottom. The red (dotted) line is the case with no marginal operators 
included. Note that at the point of instability, $p$ becomes larger than $\rho$ 
and the theory is non-causal. The blue (solid) lines are the case with marginal 
operators included. The theory is causal at all times, with $p < \rho$.}
\end{figure}

\begin{figure}
\psfig{figure=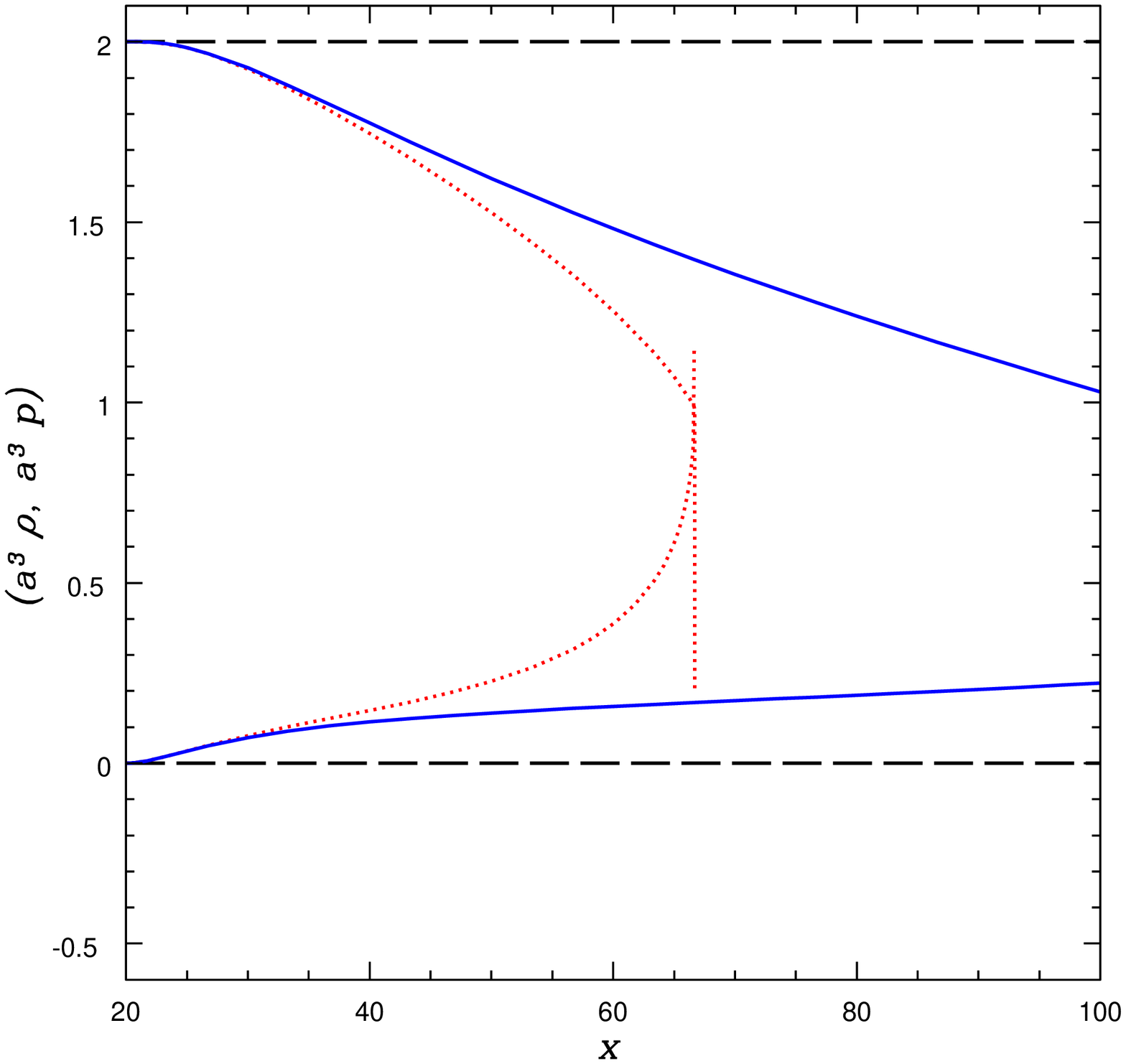}
\bigskip
\bigskip
\bigskip
\bigskip
\bigskip
\bigskip
\caption{Energy density and pressure for the case $\gamma = 0.8$. The black 
(dashed) line is the free field case: $a^3 \rho = 2$ is the top line, and $a^3 p 
= 0$ is the bottom. The red (dotted) line is the case with no marginal operators 
included.  The blue (solid) lines are the case with marginal operators included. 
This case has the same causality properties as the case with $\gamma = 5$.}
\end{figure}


\begin{references}
\bibitem{eotovos22} R. V. E\"{o}tv\"{o}s, D. Pek\'{a}r, E. Fekete, Ann. Physik 
{\bf 68}, 11 (1922).
\bibitem{will93} C. M. Will, {\it Theory and Experiment in Gravitaional Physics} 
(Cambridge University Press, Cambridge, 1993).


\bibitem{spergel99} D. N. Spergel and P. J. Steinhardt, astro-ph/9909386.
\bibitem{hannestad99} S. Hannestad, astro-ph/9912558.
\bibitem{hogan00} C. J. Hogan and J. J. Dalcanton, astro-ph/0002330.
\bibitem{burkert00} A. Burkert, astro-ph/0002409.
\bibitem{peebles00} P. J. E. Peebles, astro-ph/0002495.
\bibitem{goodman00} J. Goodman, astro-ph/0003018.
\bibitem{hannestad00} S. Hannestad and R. J. Scherrer, astro-ph/0003046.
\bibitem{hu00} W. Hu, R. Barkana, and  A. Gruzinov, astro-ph/0003365.
\bibitem{riotto00} A. Riotto and I. Tkachev, astro-ph/0003388.
\bibitem{kaplinghat00}  M. Kaplinghat, L. Knox, and M. S. Turner, 
astro-ph/0005210.

\bibitem{frieman91} J. A. Frieman and B. Gradwohl, Phys. Rev. Lett {\bf 67}, 
2926 (1991).
\bibitem{gradwohl92a} B. Gradwohl and J. A. Frieman, Astrophys. J. {\bf 398}, 407, (1992).
\bibitem{gradwohl92b} J. M. Gelb, B. Gradwohl, and J. A. Frieman, Astrophys. J. 
Lett. {\bf 403}, L5 (1993), hep-ph/9208239.
\bibitem{kinney00} W. H. Kinney and M. Brisudova, in preparation.


\bibitem{zlatev99} I. Zlatev and P. J. Steinhardt, astro-ph/9906481.
\bibitem{bento00} M. C. Bento and O. Bertolami, astro-ph/0003350.
\bibitem{matos00a}  T. Matos and L. A. Uren\~a-L\'opez, astro-ph/0004332.
\bibitem{matos00b} T. Matos and L. A. Uren\~a-L\'opez, astro-ph/0006024.
\bibitem{hu99} W. Hu and P. J. E. Peebles, astro-ph/9910222.
\bibitem{sahni99} V. Sahni and L. Wang, astro-ph/9910097.
\bibitem{matos99} T. Matos, F. S. Guzman, and L. A. Uren\~a-L\'opez, astro-ph/9908152.

\bibitem{boyanovsky94} D. Boyanovsky, H. J. de Vega, R. Holman, D. -S. Lee, and 
A. Singh, Phys. Rev. D {\bf 51}, 4419 (1995), hep-ph/9408214.
\bibitem{boyanovsky97} D. Boyanovsky, H. J. de Vega, R. Holman, A. Singh, and M. 
Srednicki, Phys. Rev. D {\bf 56}, 1939 (1997), hep-ph/9703327.
\bibitem{kofman97} For a review, see L. Kofman, A. Linde, and A. A. Starobinsky, 
Phys. Rev. D {\bf 56}, 3258 (1997), hep-ph/9704452.
\end{references}
\end{document}